\documentclass[footinbib,a4paper,aps,reprint,twocolumn,preprintnumbers,amsmath,amssymb,nobalancelastpage,longbibliography]{revtex4-2}

\usepackage[english]{babel}
\usepackage[dvipsnames]{xcolor}
\usepackage{comment}
\usepackage{graphicx}
\usepackage{stackengine}
\usepackage{dcolumn}
\usepackage{bm}
\usepackage[utf8]{inputenc}
\usepackage{amssymb}
\usepackage{amsmath}
\usepackage{pstricks}
\usepackage{slashed}
\usepackage{braket}
\usepackage{amsfonts}
\usepackage{bbm}
\usepackage{physics}

\usepackage{natbib}
\usepackage{verbatim}
\usepackage{siunitx}
\usepackage{lipsum}
\usepackage{slashed}
\usepackage{ulem}
\usepackage[left=1.85cm,right=1.85cm,top=1.85cm,bottom=1.85cm]{geometry}
\usepackage{verbatim}
\usepackage{xcolor}
\usepackage{hyperref}
\hypersetup{    colorlinks,    linkcolor={blue!50!black},    citecolor={blue!50!black},    urlcolor={blue!80!black}}

\makeatletter
\renewcommand{\p@paragraph}{} 
\makeatother

\def\zn{\mathbb{Z}_N}

\begin{document}

\title{
Field digitization scaling
in a $\mathbb{Z}_N \subset U(1)$ symmetric model
}

\author{Gabriele Calliari}\email{Gabriele.Calliari@uibk.ac.at}
\author{Robert Ott}\thanks{}
\author{Hannes Pichler}
\author{Torsten V. Zache}
\email{Torsten.Zache@uibk.ac.at}

\affiliation{Institute for Theoretical Physics, University of Innsbruck, Innsbruck, 6020, Austria}
\affiliation{Institute for Quantum Optics and Quantum Information, Austrian Academy of Sciences, Innsbruck, 6020, Austria}

\begin{abstract}
    The simulation of quantum field theories, both classical and quantum, requires regularization of infinitely many degrees of freedom. 
    However, in the context of field digitization (FD) -- a truncation of the local fields to $N$ discrete values -- a comprehensive framework to obtain continuum results is currently missing.
    Here, we propose to analyze FD by interpreting the parameter $N$ as a coupling in the renormalization group (RG) sense.
    As a first example, we investigate the two-dimensional classical $N$-state clock model as a $\mathbb{Z}_N$ FD of the $U(1)$-symmetric $XY$-model.
    Using effective field theory, we employ the RG to derive generalized scaling hypotheses involving the FD parameter $N$, which allows us to relate data obtained for different $N$-regularized models in a procedure that we term \emph{field digitization scaling} (FDS).
    Using numerical tensor-network calculations at finite bond dimension $\chi$, we further uncover an unconventional universal crossover around a low-temperature phase transition induced by finite $N$, demonstrating that FDS can be extended to describe the interplay of $\chi$ and $N$. 
    Finally, we analytically prove that our calculations for the 2D classical-statistical $\mathbb{Z}_N$ clock model are directly related to the quantum physics in the ground state of a (2+1)D $\mathbb{Z}_N$ lattice gauge theory which serves as a FD of compact quantum electrodynamics.
    Our study thus paves the way for applications of FDS to quantum simulations of more complex models in higher spatial dimensions, where it could serve as a tool to analyze the continuum limit of digitized quantum field theories.
\end{abstract}

\maketitle
\graphicspath{{./figures/}}

\paragraph*{Introduction.---}

Quantum field theories (QFTs) are powerful theoretical descriptions of nature at all energy scales, ranging from condensed matter~\cite{sachdev2023quantum} to high-energy physics~\cite{weinberg1995quantum}.
Solving QFTs is a long-standing, ongoing effort and has motivated developing extensive theoretical and numerical computing tools for classical simulations, e.g., with Monte Carlo sampling~\cite{assaad2008worldline,montvay1994quantum} or tensor networks~\cite{orus2019tensor,cirac2021matrix}, and more recently, quantum simulations using quantum hardware~\cite{bauer2023quantum,dimeglio2024quantum}. 
A common challenge in all of these approaches is the regularization of the QFT's infinitely many continuous degrees of freedom at a definite length (or energy) scale, and the task of obtaining regularization-independent results in the continuum limit.

\begin{figure}[t!]
    \centering
    \includegraphics[width=\linewidth]{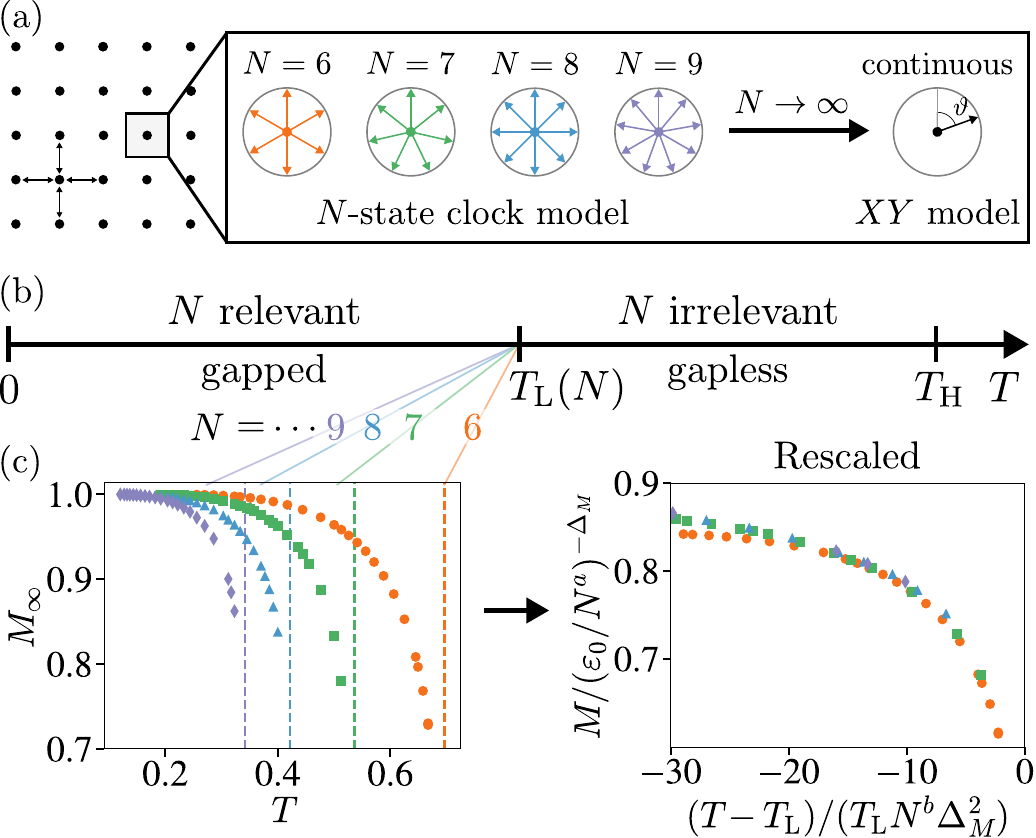}
    \caption{\textbf{Field digitization scaling in the $N$-state clock model}. (a) We investigate the $N$-state clock model with nearest-neighbor-interacting, discrete angles $\vartheta=2\pi n/N$, ($n=0,\dots, N-1$) on a 2D square lattice. In the limit $N\rightarrow \infty$, these $\zn$-symmetric models give rise to the $U(1)$-invariant $XY$-model, which is gapless for $T < T_{\rm H}$. (b) Field digitization to $N$ discrete angles is relevant only at low temperatures, leading to an ordered, gapped phase at $T\leq T_{\rm L}(N)$. (c) In this regime we uncover a self-similar scaling behavior among different $N$-regularized models. (Left) Magnetization $M$ for different~$N$ versus temperature~$T$. Dashed colored vertical lines represent $T_{\rm L}(N)$. (Right) Based on the behavior of the correlation length $\xi(T, N)$ [see Eq.~\eqref{eq:corr_infty_num} and Fig.~\ref{fig:Fig2}], we formulate an $N$-dependent scaling Ansatz [Eq.~\eqref{eq:Ansatz_N}] for local observables. Upon rescaling both axes accordingly, we observe a collapse of $M$ for different $N$ and $T$. Here, $a=1.5, b=1, \Delta_M(N)=2/N^2$.}
    \label{fig:Fig1}
\end{figure}

Most prominently, quantum fields are discretized on finite lattices~\cite{rothe2012lattice,montvay1994quantum}, where thermodynamic and continuum results can be retrieved with a finite-size scaling analysis~\cite{cardy1996scaling}.
However, in addition to finite lattices, several simulation techniques --- from quantum-inspired classical methods, such as tensor networks, to future large-scale quantum computation --- typically require a digitization of the fields, which are locally truncated to finitely many values~\cite{jordan2012quantum,klco2019digitization,delcamp2020computing,meurice2022tensor}. Removing this digitization to obtain physical continuum results, however, is challenging and further complicated in the presence of continuous symmetries, as is the case for lattice gauge theories (LGTs). 
While recently several such field-digitization strategies have been developed -- for example based on finite subgroups~\cite{bender2018digital,alexandru2019gluon,hackett2019digitizing,gonzalez-cuadra2022hardware,lamm2024block,gustafson2024primitive}, fixed representation cutoffs~\cite{byrnes2006simulating,zohar2015formulation,davoudi2021search,ciavarella2021trailhead,rhodes2024exponential,ebner2024entanglement}, orbifold constructions~\cite{buser2021quantum,bergner2024toward}, quantum groups~\cite{zache2023quantum,hayata2023string,hayata2025floquet}, large-$N_c$ expansions~\cite{ciavarella2024quantum}, quantum link models or large spins~\cite{chandrasekharan1997quantum,wiese2022quantum,kasper2016schwinger,zache2022toward,halimeh2024spin,calliari2025quantum,joshi2025efficient}, qubit regularization~\cite{alexandru2019models,singh2019qubit,bhattacharya2021qubit,maiti2024asymptotic,alexandru2024fuzzy}, or Hamiltonian truncation~\cite{ingoldby2024enhancing} 
 -- a generally applicable framework for obtaining continuum limits, and a comprehensive understanding of these different procedures in terms of the renormalization group (RG) are still missing.

In this letter, we take first steps towards an RG framework incorporating the truncation of the local field to a finite number ($N$) of values. Focusing on the example of digitizing the continuous group $U(1)$ to a finite subgroup $\zn$, we show that it is possible to interpret the truncation parameter $N$ as a coupling constant, and thus distinguish physical scenarios in terms of relevant or irrelevant field truncation. We probe this framework by systematically varying $N$ in a scaling analysis: the \textit{field digitization scaling}~(FDS).
We first demonstrate its applicability with tensor-network simulations of a two-dimensional classical $\zn$-symmetric clock model, a regularization of the continuous-field $XY$-model to $N$ discrete angles~[see Fig.~\ref{fig:Fig1}(a)]\footnote{Recently, $N$-state clock models have been the focus of several numerical studies, in particular Refs.~\cite{chen2022monte,li2020critical,li2022tensornetwork}, aiming at a precise numerical extrapolation of the critical temperatures, as well as the investigation of a possible self-dual point and the compactification radius of the conformal field theory. In contrast, in this letter we focus on \emph{relating} the $\mathbb{Z}_N$ models across distinct $N$, interpreting them as different digitizations of the \emph{same} $U(1)$ critical properties in the vicinity of the low-temperature phase transitions.}. By the quantum-classical correspondence, our results also translate to the 1D quantum $N$-state clock model~\cite{ortiz2012dualities}.
We show that for large enough $N$, rescaling the numerical data as a function of $N$ and temperature $T$ relates different $N$ regularizations in an RG sense and allows us to retrieve results that are controlled by the continuous symmetry group $U(1)$.
Furthermore, we extend our analysis to take into account the bond dimension $\chi$ of the tensor network, which plays the role of a second regulator. This motivates a novel generalized scaling Ansatz of $N, T$ and $\chi$, with which we uncover an unconventional crossover regime.

As a generalization, we demonstrate that our numerical results have direct consequences for quantum gauge theories. Specifically, we prove an explicit relation between our chosen tensor-network representation and the ground state of a (2+1)D $\zn$-symmetric LGT, which provides an asymptotic regularization of compact quantum electrodynamics.
This indicates that our results apply to a much larger class of classical and quantum models, and our FDS analysis thus paves the way towards a more complete understanding of how to obtain continuum results from field-digitized models.
We anticipate that such an approach will help to assess digitization errors and to estimate resources required in the simulation of QFTs at different truncations~\cite{meth2025simulating,dandrea2024new,paulson2021simulating,haase2021resource}, ultimately identifying optimal digitization strategies and thus enabling more efficient use of quantum simulations.

\paragraph*{Field digitization.---}As an example of a field-digitized model, we consider the $N$-state clock model (with $N>5$~\footnote{In this work, we are interested in the behavior $N\geq 5$, since the detailed universal aspects of $N$-state clock models differ for small $N< 5$~\cite{li2020critical}.}), described by the Hamiltonian
\begin{equation}
  H_{N}=-J \sum_{\langle i j\rangle} \cos(\vartheta_i - \vartheta_j)\;, \quad \vartheta_i=\frac{2\pi n_i}{N},
 \label{eq:clock_Ham}
\end{equation}
with $n_i=0, \dots, N-1$, and $J=1$ the coupling strength. Here, the degrees of freedom $\vartheta_i$ are digitized, $\zn \subset U(1)$ classical angles, interacting with their nearest-neighbors on a two-dimensional square lattice [see Fig.~\ref{fig:Fig1}(a)].
This model has been extensively studied in the literature for various $N$~\cite{elitzur1979phase,ueda2020finite,chatelain2014dmrg}, and in the limit $N\rightarrow \infty$ it reduces to the $XY$-model~\cite{berezinskii1971destruction,berezinskii1972destruction,kosterlitz1974critical}. 
The partition function $Z(N, \beta)$ for an infinite system, where $\beta=1/T$ is the inverse temperature, can be directly encoded in a 2D tensor network (see End Matter (EM)~\ref{sec:app_state_construction})
\begin{align}
  Z(N, \beta)=\sum_{\{\vartheta_i\}}e^{-\beta H_N}=&\vcenter{\hbox{\includegraphics[height=1.4cm]{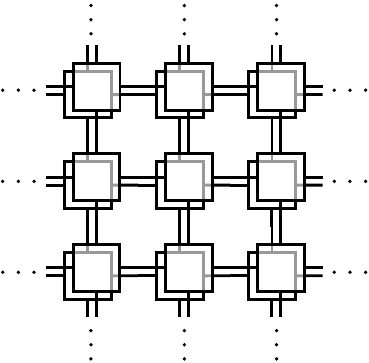}}}\xleftarrow{\chi\rightarrow \infty} \vcenter{\hbox{\includegraphics[height=1.4cm]{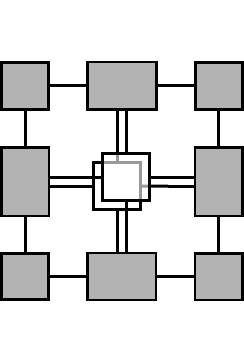}}} \;.
  \label{eq:partition_clock}
\end{align}
As indicated on the right, we further regulate the infinite system with truncated environment tensors, representing the partition function $Z(N,\beta, \chi)$, which we obtain through the isotropic corner transfer matrix renormalization group (CTMRG) algorithm \cite{fishman2018faster}. This introduces the bond dimension $\chi$, which, similar to a finite system size, truncates the correlations of the system. 
The full partition function is recovered as $Z(N, \beta, \chi)\xrightarrow{\chi\rightarrow \infty} Z(N,\beta)$.

\begin{figure}[t!]
    \centering
    \includegraphics[width=\linewidth]{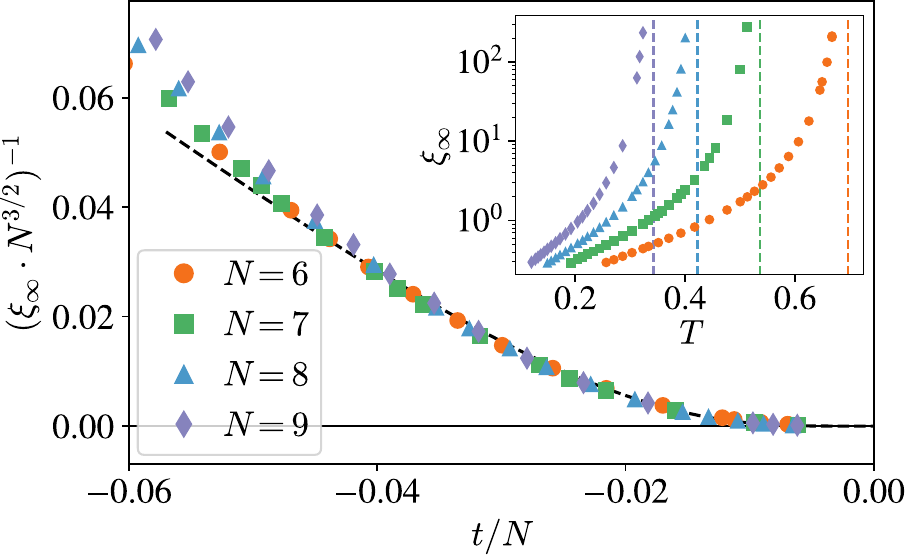}
        \caption{\textbf{Relevant field digitization and universal behavior}. In the gapped, ordered phase we uncover a universal scaling form of the correlation length. Shown is the extrapolated correlation length $\xi_\infty=\xi(\chi\rightarrow \infty)$ as a function of the reduced temperature $t=(T-T_{\rm L})/T_{\rm L}$ for field dimensions $N=6,7,8,9$. Data points collapse on a single curve [dashed black line, see Eq.~\eqref{eq:corr_infty_num}] upon rescaling with~$N$. (Inset) Unrescaled correlation length $\xi_\infty$ versus temperature $T$. Colored vertical dashed lines represent~$T_\mathrm{L}(N)$.}
    \label{fig:Fig2}
\end{figure}

To interpret the impact of the truncation parameter $N$ from an RG perspective, we use an effective description in terms of continuous fields $\theta(x), \phi(x)$, with $x=(x_1, x_2)$ coordinates in two-dimensional Euclidean space. The field $\theta$ is the continuous-variable generalization of the microscopic variable $\vartheta$, and the dual field $\phi$ obeys $\partial_\mu \phi=-\varepsilon_{\mu\nu} K \partial_\nu \theta$, where $\partial_\mu=\partial/\partial x_\mu ,\mu=1,2$, and $K=K(T)$ is the Luttinger parameter. One can show that the expected low-energy effective action for the model $H_N$ takes the form~\cite{wiegmann1978onedimensional,li2020critical}
\begin{align}
     &S_N=S_{\rm sG}+ \frac{g_\mathrm{L}}{2\pi \alpha^2}  \int{\rm d}^2x \; \cos{(N\sqrt2  \theta)},      \label{eq:clock_action}
 \\
      &S_{\rm sG}=\int{\rm d}^2x \;\bigg\{ \frac{1}{2\pi K} (\nabla\phi)^2 + \frac{g_\mathrm{H}}{2\pi \alpha^2} \cos{(\sqrt2  \phi)}\bigg\}\, .
      \label{eq:sineGordon_action}
\end{align}
$S_{\rm sG}$ describes the Euclidean sine-Gordon (sG) field theory, with coupling constant $g_{\rm H}=g_{\rm H} (T, N)$ and UV-cutoff $\alpha$, while $g_{\rm L}=g_{\rm L}(T,N)$ is the coupling constant of a dual cosine interaction generated by the truncation. The latter term explicitly depends on $N$, revealing that $N$ plays the role of a coupling parameter.
At temperatures $T> T_{\rm L}$ the additional operator $\sim \cos(N\sqrt{2}\theta)$ is RG-irrelevant [see Fig.~\ref{fig:Fig1}(b)], and we recover the standard sG theory, with a critical, gapless phase below~$T_{\rm H}$ \footnote{In this work, we do not investigate the second BKT transition at $T=T_{\rm H}$, whose behavior is essentially independent of~$N$}.
Conversely, for $T<T_{\rm L}(<T_{\rm H})$, $N$ induces a relevant perturbation, leading to an ordered, gapped phase, which is separated from the gapless phase by a Berezinskii-Kosterlitz-Thouless (BKT) transition at $T_{\rm L}(N)$~\cite{ortiz2012dualities,li2022tensornetwork}.

\paragraph*{Field digitization as a relevant perturbation.---}

To demonstrate that the field digitization $N$ can be interpreted as a relevant perturbation, we consider the correlation length $\xi$ next. Specifically, we focus on the ordered phase $T<T_{\rm L}$ in the vicinity of the critical point (CP), and investigate how the field digitization $N$ affects the observables.

In particular, we relate the functional behavior of $\xi$ to an RG analysis involving $N$. Below $T_H$ the operator $\cos{(\sqrt{2}\phi)}$ is irrelevant, and $S_N$ effectively reduces to a dual sG action.
From the corresponding RG flow equations~\cite{kehrein2001flow}, we derive that close to the CP the correlation length scales as $\xi \propto \exp\{\pi/(4 \cdot \sqrt{C}) \}$, with $C\approx (g_{\rm L}(T,N)/2\pi)^2$ (see EM~\ref{sec:app_RG_analytic}). 
We then explicitly determine the $N$-dependence in $C$ and in the prefactor from our numerical simulations. 
Since the ordered phase is gapped, we can extrapolate the data points to the limit $\chi~\rightarrow~\infty$; our numerical results for $\xi_\infty=\xi(\chi\rightarrow \infty)$ indeed confirm the derived scaling with
\begin{equation}
    \xi_\infty(T, N)=\frac{\varepsilon_0}{N^a} \exp\left(\frac{\pi}{4}\frac{1}{\sqrt{|t|/N^b}}\right),
    \label{eq:corr_infty_num}
\end{equation}
where $a\approx 1.5, \; b\approx1,\; \varepsilon_0\approx \log(2)$ are consistent with our numerical results, while $t~=~{(T-T_{\rm L}(N))/T_{\rm L}(N)}~<~0$ labels the distance from the CP (see Fig.~\ref{fig:Fig2})\footnote{In this work we use the critical temperatures $T_{\rm L}(N)$ numerically determined in Ref.~\cite{li2022tensornetwork}.}\footnote{By comparing our results Eq.~\eqref{eq:corr_infty_num} to the derived $\xi$ behavior, we infer $C\approx |t|/N$, which can also be interpreted as an $N$-dependent UV initialization of $g_{\rm L}(T, N)\propto \sqrt{|t|/N}$.}. Therefore, our numerical simulation confirms the interpretation of the truncation parameter $N$ as an RG coupling.

\begin{figure}[t!]
    \centering
     \includegraphics[width=\linewidth]{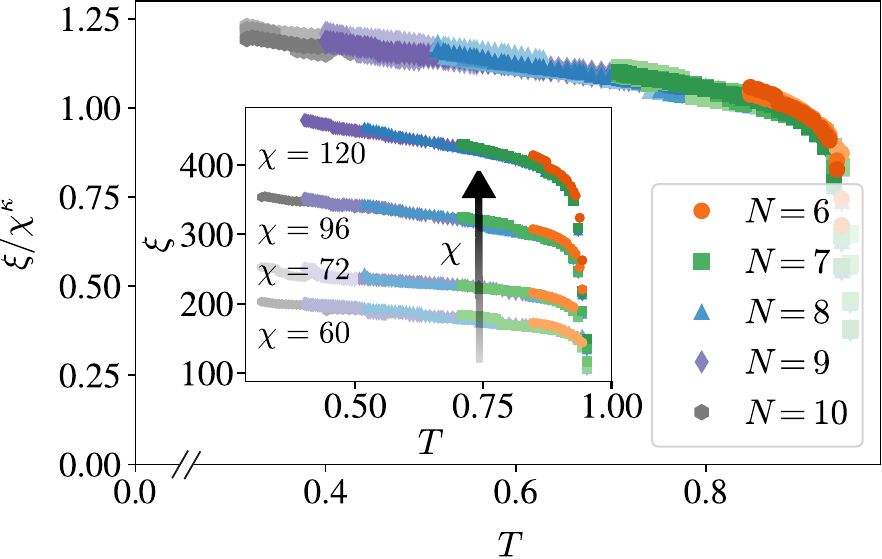}
    \caption{\textbf{Emergent $U(1)$ symmetry at finite $N$}. 
    In the gapless, critical phase the field digitization $N$ is an irrelevant perturbation: (Inset) the trivial collapse in $N$ reveals the emergent $U(1)$ symmetry. Instead, the finite $\chi$ truncates the otherwise diverging correlation length $\xi$, shown versus $T$ for different $N$ and increasing bond dimension $\chi$ (increasing intensity). 
    (Main) Upon rescaling the vertical axis as $\xi/\chi^\kappa$, we observe a collapse for different $\chi$ and $N$ as a function of temperature $T$. 
    }
    \label{fig:Fig3}
\end{figure}

We now use the uncovered behavior of the correlation length $\xi_\infty$ to make further predictions on the field digitization scaling of local observables. According to standard scaling arguments~\cite{cardy1996scaling}, close to the CP~$T_{\rm L}$ an observable $O_\infty(T, N)$ with scaling dimension $\Delta_O(N)$ scales as $O_\infty(T, N)\sim \xi^{-\Delta_O(N)}$. Note that here the label ``$\infty$'' indicates that the observable $O$ is independent of $\chi$, i.e., the regulator is removed. Based on Eq.~\eqref{eq:corr_infty_num}, we can then formulate the scaling Ansatz
\begin{equation}
    O_\infty(T,N) = \left( \frac{\varepsilon_0}{N^a} \right)^{-\Delta_O(N)} f\left[\frac{|t|}{N^b\cdot \Delta^2_O(N)} \right]\, ,
    \label{eq:Ansatz_N}
\end{equation}

with scaling function~$f[x]$.
We test our Ansatz for the extrapolated magnetization $M_\infty=M(\chi\rightarrow \infty)$, with $M= \cos(\vartheta_i)$, which translates to the operator $\cos(\sqrt{2}\theta)$ in the low-energy effective theory Eq.~\eqref{eq:clock_action}.
As shown in Fig.~\ref{fig:Fig1}(c), upon properly rescaling the numerical results according to Eq.~\eqref{eq:Ansatz_N} and with $\Delta_M=2/N^2$, we obtain a collapse of the data points for different $N$ and $T$, providing further evidence for our Ansatz. 
Note that, in contrast to standard scaling approaches, here the truncation parameter $N$ also appears in the exponents through an $N$-dependent scaling dimension.

Moreover, we emphasize that our approach is applicable in the neighborhood of $T_L (N)$ for finite $N$, in contrast to the simpler rescaling $T/T_{\rm L}(N)$ which is applicable close to $T=0$ for $N\rightarrow \infty$ as motivated by the mapping to a discrete Gaussian model~\cite{hasenbusch1997computing} [see Supplemental Material for a detailed comparison].

\begin{figure}[t!]
    \centering
    \includegraphics[width=\linewidth]{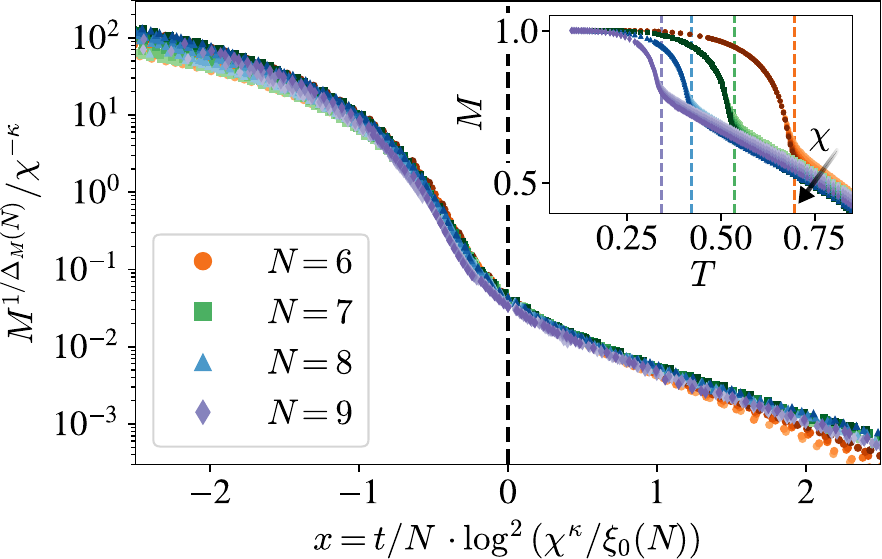}
    \caption{\textbf{Crossover regime}. Close to the CP $T=T_{\rm L}$ both truncations in $N$ and $\chi$ are relevant --- rescaling according to Eq.~\eqref{eq:scalingAnsatz} leads to a scaling collapse. (Main) Rescaled magnetization $M$ is shown as a function of the parameter $x=t/N \cdot \log^2{(\chi^\kappa/\xi_0(N))}$ for several $N$ (different colors and shapes) and bond dimension $\chi$ (increasing intensity, from $\chi=60$ up to a maximum of $\chi=192$ for $N=6,7$).
    (Inset) Unrescaled magnetization $M$ versus temperature $T$. Vertical dashed lines represent $T_\mathrm{L}(N)$. The black shaded arrow indicates increasing~$\chi$.}
    \label{fig:Fig4}
\end{figure}

\paragraph*{Irrelevant field digitization and emerging $U(1)$-symmetry.---}

Let us now consider the regime in which the perturbation $N$ is irrelevant, i.e., we focus on the gapless region ($T_{\rm L} < T < T_{\rm H}$). Deep in the critical phase, in agreement with predictions from the effective field theory, our numerical results confirm the expected collapse among different $N$--- revealing the emergence of the $U(1)$-invariance already at small values of $N$ (see Insets of Fig.~\ref{fig:Fig3} and Fig.~\ref{fig:Fig4}).
In this regime the physics is captured by the Luttinger liquid conformal field theory (CFT).
Due to the finite bond dimension~$\chi$, our tensor-network calculations introduce an additional relevant perturbation, which truncates the otherwise diverging correlation length as $\xi\propto \chi^\kappa$ (with $\kappa$ a parameter fixed by the CFT, which we numerically extrapolate to $\kappa=1.247$) \cite{tagliacozzo2008scaling,pollmann2009theory}.
In agreement with previous observations~\cite{ueda2020finite,li2020critical}, the numerical data in this gapless phase depend on $\chi$, yet upon rescaling the $y$-axis as $\xi/\chi^\kappa$, we obtain a collapse for multiple $N, \chi$ and $T$ (see Fig.~\ref{fig:Fig3}).
Close to the gapped ordered regime we observe some deviations in $\chi$ and $N$ from the universal curve, which we attribute to the impact of the operator $\cos{(N\sqrt{2}\theta)}$ and its non-trivial interplay with the truncation in $\chi$, which we study in detail next.

\paragraph*{Crossover regime.---}

So far, we have separately investigated the effects of the regularizations introduced by $N$ and $\chi$ on the two sides of the low-temperature phase transition. That is, we studied regimes sufficiently far away from the CP, where one of these perturbations dominates. Instead, we now focus on the region closer to the CP $T=T_{\rm L}$ and investigate the interplay of $N$ and $\chi$, as well as of the temperature~$T$.
To capture their combined effect, we formulate a scaling Ansatz for general local observables $O=O(T, N, \chi)$.

To start, motivated by the uncovered behavior of $\xi_\infty(T, N)$ in Eq.~\eqref{eq:corr_infty_num}, we introduce the scaling variable $x~=~t/N\cdot \log^2(\xi/\xi_0(N))$, with $\xi_0(N)=\log{(2)}/N^{3/2}$, which by definition should remain constant in the critical regime. 
As shown in the previous section, the finite $\chi$ truncates the correlation length $\xi\sim\chi^\kappa$ and the observable $O\propto \xi^{-\Delta_O(N)}\sim\chi^{-\kappa\Delta_O(N)}$ in both the gapless phase and close to the CP \footnote{Due to the truncation in $\chi$, the scaling variable $x=x(\xi)$ cannot be a constant in the whole critical region, but rather it quantifies the proximity to the CP.}.
In analogy with the scaling Ansätze for continuous phase transitions~\cite{tagliacozzo2008scaling},
we thus introduce a generalized scaling Ansatz
\begin{align}
     O(T, N, \chi)=\Big(\chi^{\kappa}g\Big[\frac{t}{N}\log^2\Big(\frac{\chi^\kappa}{\xi_0(N)}\Big)\Big] \Big)^{-\Delta_O(N)},
    \label{eq:scalingAnsatz}
\end{align}
with scaling function~$g[x]$. Note that for an $N$-independent scaling dimension $\Delta_O$, we retrieve the usual scaling Ansatz $O(T, \chi)=\chi^{-\kappa \Delta_O} h[x]$, with the scaling function  $h[x]=(g[x])^{-\Delta_O}$ \cite{ueda2020finite,pollmann2009theory,tagliacozzo2008scaling}. 

We test our scaling hypothesis with the magnetization $M$, which at the CP has an $N$-dependent scaling dimension $\Delta_M(N)=2/N^2$. As demonstrated in Fig.~\ref{fig:Fig4}, rescaling the axes according to Eq.~\eqref{eq:scalingAnsatz} leads to a scaling collapse around the CP for a wide range of values of $N$ and $\chi$, confirming the validity of our generalized Ansatz.

\paragraph*{Generalizations.---}

Until here, our investigation has focused on an example of a 2D classical statistical model.
However, the standard quantum-to-classical correspondence implies that our FDS analysis also applies to the 1D quantum clock model, where $N$ represents the local Hilbert space dimension $\dim(\mathcal H)=N$. Throughout this work, we have used the fact that classical 2D models can also be encoded in 2D tensor-network states~\cite{verstraete2006criticality} suggesting a further extension to 2D quantum models. 
We thus expect that a similar analysis could be extended to other one- and higher-dimensional quantum models, where we believe our approach to have widespread consequences for the quantum simulation of models with (local) continuous-group internal symmetries, in particular lattice gauge theories.

\begin{figure}[t!]
    \centering
    \includegraphics[width=\linewidth]{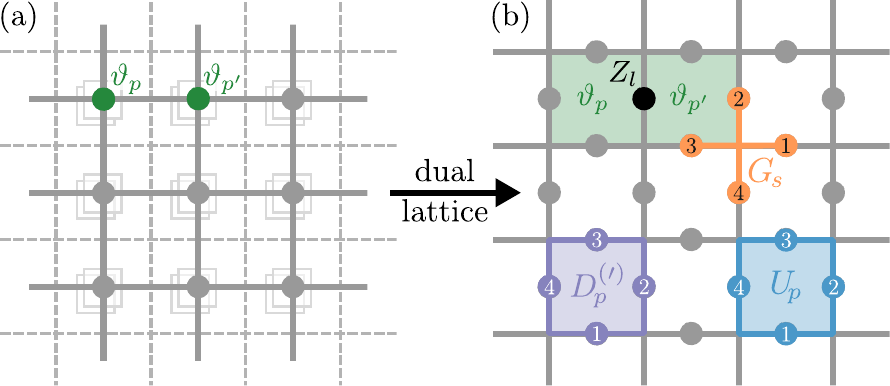}
    \caption{\textbf{Connecting the classical $N$-state clock model to a (2+1)D $\zn$ LGT}. The tensor-network encoding of the 2D classical partition function $Z_N$ [(a)] captures the quantum ground state of a $\zn$ LGT with Hamiltonian given in Eq.~\eqref{eq:LGT_Ham} on the dual lattice [(b)].
    (a) On each site $p$ of a 2D lattice we define the microscopic, discrete angle $\vartheta_p$ [see Eq.~\eqref{eq:clock_Ham}]. The partition function $Z_N$ [Eq.~\eqref{eq:partition_clock}] on this lattice can be encoded in a 2D tensor network, as depicted in the background. The dashed lines represent the dual lattice, each site of which is mapped to a plaquette of the original lattice (and vice versa). (b) On the dual lattice we define plaquette [$U_p=X_{p,1}X_{p,2}X_{p,3}^\dagger X_{p,4}^\dagger$] and Gauss' law operators [$G_s=Z_{s,1}Z_{s,2}Z_{s,3}^\dagger Z_{s,4}^\dagger$], where the labels $1,..,4$ refer to links near plaquettes~$p$ and vertices~$s$ as indicated. 
    Similarly, we define the operators $D_{p}^{(\prime)}=f(Z_{p,l})$, with $l=1,.., 4$ (see EM~\ref{sec:app_LGT}).
    The operator $Z_l$ on the link $l=\langle p, p'\rangle$ between two plaquettes $p, p'$ is defined by $\vartheta_{p}$ and $\vartheta_{p'}$ as $Z_l\equiv \exp\{i(\vartheta_{p}-\vartheta_{p'})\}$.}
    \label{fig:Fig5}
\end{figure}

As an example, we now relate the above 2D classical $N$-state clock model to a (2+1)D quantum $\zn$ LGT.
Specifically, consider the 2D quantum Hamiltonian
\begin{align}
    H_{\rm LGT}=\frac{1}{2}\sum_p  \Big[ \big(D_p(\beta) + D_p'(\beta)\big)-(U_p + U_p^\dagger)\Big],
    \label{eq:LGT_Ham}
\end{align} 
with plaquette operators $U_p=X_{p,1}X_{p,2}X_{p,3}^\dagger X_{p,4}^\dagger$ and $D_p^{(\prime)}(\beta)= \exp\{-(\beta/4)[\sum_{l=1}^4 c^{(\prime)}_l Z_{p,l} +{\rm H.c.}]\}$ (see EM~\ref{sec:app_LGT}, and for a sketch Fig.~\ref{fig:Fig5}). Here, $c_l^{(\prime)}$ are judiciously chosen, $N$-dependent complex coefficients, $\beta$ represents a coupling constant, which we eventually identify as the inverse temperature of the 2D classical model, while $Z$ and $X$ are $N$-dimensional clock and shift operators. Moreover, we restrict ourselves to gauge-invariant states, that is, we impose the Gauss' law constraints $G_s=Z_{s,1}Z_{s,2}Z_{s,3}^\dagger Z_{s,4}^\dagger~=~1$ on all vertices $s$. 
In the limit $\beta\ll 1$, the operators $(D_p(\beta)+D_p'(\beta))/2$ simplify to $- \beta(1-\cos(2\pi/N)) \sum_{l=1}^4 (Z_{p,l}+Z_{p,l}^\dagger)$, such that the Hamiltonian reduces to a standard $\zn$ LGT~\cite{horn1979hamiltonian}, that we interpret as a field digitization of (2+1)D compact quantum electrodynamics.

In EM~\ref{sec:app_LGT}, extending an analytical result from the $N=2$ case~\cite{castelnovo2008quantum} to arbitrary $N$, we prove that the ground state of the $\zn$ LGT [Eq.~\eqref{eq:LGT_Ham}]\footnote{For simplicity, here we state the result for an infinite plane, where the ground state is unique.} is given by
\begin{equation}
    \ket{\vartheta}=\sum_{\{\vartheta_p\}} e^{\frac{\beta}{2}\sum_{\langle p, p'\rangle} \cos{(\vartheta_p - \vartheta_{p'})}} g(\{\vartheta_p\})\ket{\Omega},
    \label{eq:gs_LGT}
\end{equation}
with $Z_\vartheta=\braket{\vartheta}{\vartheta}=\sum_{\{\vartheta_p\}} e^{\beta \sum_{\langle p, p'\rangle}\cos{(\vartheta_p - \vartheta_{p'})}}$ being the corresponding partition function. Here, $\ket{\Omega}=\otimes_l \ket{0}$ is a trivial product state on the dual lattice, and $\vartheta_p$ are digitized, angular variables defined on sites ~$p$, associated with the plaquettes of the dual lattice. Applying the operators $g(\{\vartheta_p\})=\prod_p (U_p)^{n_p} $, with $\vartheta_p=2\pi (n_p \mod N)/N$, on $\ket{\Omega}$ creates all possible gauge-invariant configurations.
The ground state $\ket{\vartheta}$ can be directly encoded in a tensor network of finite bond dimension, which corresponds to the tensor network in Eq.~\eqref{eq:partition_clock}, i.e., $Z_\vartheta=Z(N, \beta)$ (see also EM~\ref{sec:app_state_construction})~\cite{verstraete2006criticality}.
This identification also applies to all (equal-time) correlation functions of diagonal operators, and hence our FDS results translate to the (2+1)D $\zn$ LGT.

\paragraph*{Conclusion \& Outlook.---}

In this work, we introduced \emph{field digitization scaling} as an extension of the standard scaling paradigm by relating theories with different truncations in the number of field states. We thus take first steps towards a comprehensive RG framework for field-regularized models, with the ultimate goal of controlling continuum limits in corresponding classical and quantum simulations.

We initiate this program at the hand of numerical tensor-network calculations of one of the best studied 2D classical-statistical models, and, while some of the here presented results are specific to this case (e.g., the BKT nature of the critical point $T_{\rm L}$), we expect our results to be relevant for a wider range of models also in higher dimensions. In this context, an immediate follow-up question addresses FDS in $(2+1)$D $N$-state quantum clock models, where emergent $U(1)$ symmetries are expected and display an intriguing interplay with dangerously irrelevant clock perturbations~\cite{patil2021unconventional,shao2016quantum}. Furthermore, FDS may provide a useful perspective on the interplay between different relevant perturbations more generally, e.g., for those associated with finite system size and bond dimension.

Beyond fundamental interest, our results have practical implications for quantum simulations of QFTs -- where we are especially interested in the case of lattice gauge theories -- as a digitization of the local Hilbert space in presence of continuous symmetry groups poses an outstanding challenge.
In this context, a detailed understanding of FDS may thus open the door for quantitative resource estimates for reaching the continuum limit with quantum hardware.
For example, in the present $N$-state clock model the resources necessary to simulate critical properties at a given $T$ and $N$ can be estimated by lower-bounding the system size~$L$ of a 1D quantum clock chain by the correlation length $\xi(T, N)$ in Eq.~\eqref{eq:corr_infty_num}. This results in the value $N=6$, corresponding to the smallest total Hilbert space dimension~$\sim N^L$, as the most economical choice. It would be interesting to perform similar analyses for two-dimensional LGT models where the extent of the desired confined phase is expected to grow upon increasing the digitization parameter $N$, suggesting a non-trivial tradeoff between system size and local Hilbert space dimension.

\paragraph*{Acknowledgements.---}
This work is supported by the
European Union’s Horizon Europe research and innovation program under Grant Agreement No.~101113690
(PASQuanS2.1), the ERC Starting grant QARA (Grant No.~101041435), the EU-QUANTERA project TNiSQ (N-6001), the Austrian Science Fund (FWF) (Grant No. DOI
10.55776/COE1).
The computational results presented here have been achieved using the LEO HPC infrastructure of the University of Innsbruck.

\paragraph*{Data availability.---} The data that support the findings of this article are openly available~\cite{zenodo}.

\bibliographystyle{apsrev4-1} 
\bibliography{bibliography}

\clearpage
\appendix

\setcounter{equation}{0}
\setcounter{table}{0}

\makeatletter
\renewcommand*{\bibnumfmt}[1]{[S#1]}
\setcounter{subsection}{0}
\renewcommand{\theparagraph}{Appendix~\Alph{paragraph}}
\renewcommand{\theparagraph}{\Alph{paragraph}}

\onecolumngrid
\section*{ END MATTER }
\twocolumngrid

\paragraph{Tensor-network simulations.---}
\label{sec:app_state_construction}

In this section we provide details on our numerical simulations.
We numerically investigate the 2D classical $N$-state clock model [Eq.~\eqref{eq:clock_Ham}] with an infinite 2D tensor network by encoding the partition function $Z$ in an infinite projected entangled-pair states (iPEPS) state $\ket{\psi}$~\cite{verstraete2006criticality}. 

To derive an explicit form of $\ket{\psi}$, we start from a product state $\otimes_i\ket{+}$, with $\ket{+}=\sum_{n=0}^{N-1}\ket{n}$, and then apply the operator $\exp{-\beta h_{ij}/2}$ on all bonds between two nearest-neighbor tensors on sites $i,j $, with $\beta$ the inverse temperature. Here, $h_{ij}~=~-(Z_i Z_j^\dagger + Z^\dagger_i Z_j)/2$ are local (two-site) contributions to the Hamiltonian of the $N$-state clock model $H_N=\sum_{\langle ij \rangle} h_{ij}$, where $Z_i~=~e^{i\vartheta_i}$, with $\vartheta_i=2\pi n_i/N$ ($n_i=0, \dots, N-1$), and $J=1$.
Using $Z_i^N=1$, we decompose the exponential as 
\begin{align}
\textstyle
    e^{-\frac{\beta}{2} h_{ij}}=
\vcenter{\hbox{\includegraphics[height=1cm]{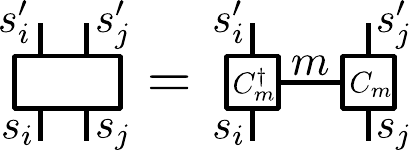}}}\, ,
\label{eq:app_gatedecomposition}
\end{align}
where the tensors on the RHS correspond to $(C_m)_{{s_i}{s_i'}}=\sqrt{c_m(\beta/2)}(Z^{ m})_{{s_i}{s_i'}}$ and its Hermitian conjugate $C_m^\dagger$,
with real coefficients 
\begin{equation}
    c_m\left(\frac{\beta}{2}\right)=\frac{1}{N}\sum_{k=0}^{N-1} \cos{\left(\frac{2\pi km}{N}\right)}e^{-\frac{\beta}{2} \cos{\left(\frac{2\pi k }{N}\right)}}\,.
\end{equation} 
Here the labels $s_i, s_i'$ indicate the matrix element $(s_i, s_i')$ in the matrix representation of the operator (see Appendix~\ref{sec:app_LGT}).
We thus define the local PEPS tensor $A_\psi$ for $\ket{\psi}$ by applying the tensors $C_m^{(\dagger)}$ to the (product state) PEPS $A_+$ that corresponds to $\ket{+}$, as shown in Fig.~\ref{fig:App_TN_encoding}(a). 

Due to the asymmetry between the tensors $C_m^{(\dagger)}$ in the decomposition Eq.~\eqref{eq:app_gatedecomposition}, $A_\psi$ is not symmetric under spatial rotations, in apparent conflict with the rotational invariance of $H_N$. It is useful to restore this symmetry as follows.
We first rewrite
$C_m^{\dagger}=\sqrt{c_m}(Z^\dagger)^m=U_{N-m,m'} \cdot \sqrt{c_{m'}}Z^{m'}$, with the unitary $U_{m,m'}=\delta_{m, N-m'}$, which can be further decomposed as $U=O\cdot O^T$. This allows us to gauge the PEPS according to 
\begin{align}
\vcenter{\hbox{\includegraphics[width=0.84\linewidth]{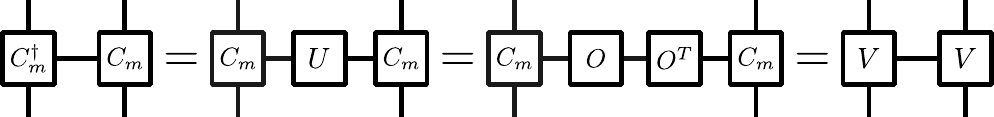}}} \;.
\end{align}
We thus obtain the operator $V$, which corresponds to a symmetric decomposition of $\exp\{-\beta h_{ij}/2\}$.
By applying $V$ on the tensor $A_+$, as shown in Fig.~\ref{fig:App_TN_encoding}(b), we obtain the desired rotational-invariant representation $B_\psi$ of $\ket{\psi}$.
\setcounter{figure}{5}
\begin{figure}[h!]
    \centering
    \includegraphics[width=0.8\linewidth]{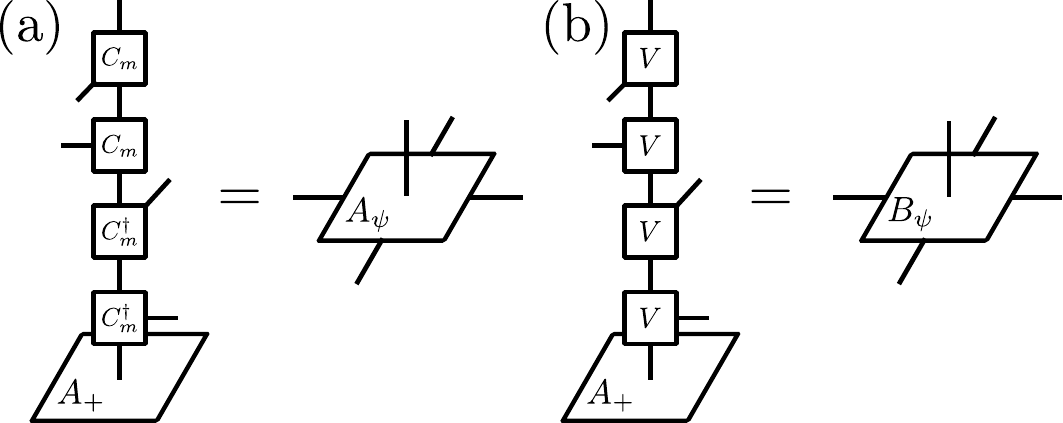}
    \caption{\textbf{Tensor-network encoding of the state $\ket{\psi}$}. (a) The tensor $A_\psi$ is obtained by applying on $A_+$ (representing $\ket{+}$) the operators $C_m^{(\dagger)}=\sqrt{c_m}Z^{(\dagger) \,m}$ in two spatial directions each. (b) A rotational-invariant encoding $B_\psi$ is instead obtained by symmetrically applying the tensors $V$. }
    \label{fig:App_TN_encoding}
\end{figure}    

Since we mainly focus on local observables, we can restrict ourselves to a single site and approximate the contraction of the infinite tensor network with environment tensors (ETs): given the rotational symmetry, we only use a single corner transfer matrix $C$ and one half-row transfer matrix $T$, which we optimize using the isotropic CTMRG algorithm~\cite{fishman2018faster,naumann2024introduction}\footnote{Our standard implementation of the CTMRG algorithm closely follows Ref.~\cite{naumann2024introduction}, and our isotropic code is inspired by Ref.~\cite{fishman2018faster}.}.

In our numerics we further impose ferromagnetic boundary conditions as follows. We first run CTMRG for increasingly small bias magnetic fields $h\neq 0$, obtained by adding to $H_N$ the term $h(Z_i+Z_i^\dagger)$ to bias the symmetry-broken phase towards the angle along $\vartheta=0$. We then use the resulting ETs as initial Ansätze for the successive optimizations at smaller $h$. Specifically, we compute ETs for $h=10^{-2}, 10^{-4}$ with a convergence parameter $\epsilon=10^{-5}$, where $\epsilon$ is the maximal difference between the values of the (diagonal) matrix $C$ in two subsequent steps.
In the end, we run one final CTMRG with initial ETs optimized at the smallest $h\neq0$ for the $h=0$ case with $\epsilon=10^{-8}$.

Having determined both $A_\psi$ and the ETs $C, T$, we evaluate local observables. The magnetization $M$ is given by the single-site expectation value $M=\langle Z+Z^\dagger\rangle/2$. By contracting $T$ with itself we also build a row transfer matrix from which we compute the correlation length $\xi=1/\ln(\lambda_1/\lambda_2)$, with $\lambda_1, \lambda_2$ the two largest eigenvalues. 
In our numerical analysis we discarded clearly deviating data points.

\paragraph{Functional form of the correlation length $\xi$ from the RG flow equations.---}
\label{sec:app_RG_analytic}

In this section we elaborate on the RG calculations used to derive the functional form of the correlation length $\xi$.
In the action $S_N$ Eq.~\eqref{eq:clock_action} 
for $T<T_{\rm H}$ the term $\cos(\sqrt{2}\phi)$ becomes irrelevant, thus allowing us to map $S_N$ to the standard sG action \cite{kehrein2001flow} 
\begin{equation}
    S_\mathrm{sG} = \int {\rm d}^2x\; \left\{\frac{1}{2}(\partial_\mu \varphi)^2 + \frac{u}{\alpha^2} \cos{(\beta_{\rm sG} \varphi)}\right\},
\end{equation}
using the substitution $N \theta \sqrt{2} \rightarrow \beta_{\rm sG} \varphi$, 
as well as the parameter identifications $\beta_{\rm sG}^2 = 2\pi N^2/K$, and $u=g_{\rm L}(T,N)/2\pi$.

Close to the CP $\beta_{\rm sG}^2~\approx~8\pi$, the sG RG flow equations are given by $
{\rm d}\tilde \beta^2/{\rm d} \ln\Lambda=4 u^2, \; {\rm d} u/{\rm d}\ln\Lambda=(\tilde \beta^2 -2 ) u$, 
with $\tilde \beta^2 =\beta^2_\mathrm{sG}/4\pi$, while $\Lambda=e^{-l}$ defines the (inverse) length scale~\cite{kehrein2001flow}. Furthermore, at the CP the Luttinger parameter $K_{\rm L}(N)=N^2/4$, and using the expansion $K=K_{\rm L}+\delta K$, we find $d=1-\tilde \beta^2/2=4\delta K/N^2$, with $0<d \ll 1$ in the gapped phase close to the CP. We further assume $u>0$. 
Rewriting the RG equations in terms of the variables $z_{1/2}=u \pm d$, as ${\rm d} z_{1/2}/{{\rm d} \ln\Lambda}= \mp z_{1/2} (z_1 + z_2)$ and
using that $z_1\cdot z_2=u^2 -d^2=C(N,T)$ is a constant of the flow, 
we obtain $
   {\rm d} z_1/{\rm d} \ln\Lambda = -(z_1^2 + C)$. The solution is $
 z_1(l)=\sqrt{C}\tan{(\sqrt{C} l +c_1)} $,
with $c_1$ a constant.
We can then rewrite $u(l)~=~(z_1+C/z_1)/2$, and fix $c_1=\arctan{\{(\varepsilon+\sqrt{\varepsilon^2-C}/C)/\sqrt{C}\}}$ from the initial value $u(0)=\varepsilon$. At the start of the RG flow, both $u(0), d(0)\ll 1$; due to the closeness to the CP we further assume $1\gg u(0)\gg d(0)>0$, i.e., $ C\approx \varepsilon\ll 1$. Hence, $c_1\approx \arctan{(\varepsilon/\sqrt{C})}\approx \pi/4$.
Consider the flow parameter $l^*$ at which $u(l^*)=1$, which equals
\begin{equation}
    \textstyle
    l^*=\frac{1}{\sqrt{C}}\Big(-c_1+\arctan{\Big[\frac{1+\sqrt{1-C}}{\sqrt{C}}\Big]}\Big)\approx\frac{\pi}{4\sqrt{C}}.
\end{equation}
This identifies the characteristic correlation length as $\xi=e^{l^*}$, equal to $
 \xi \propto \exp\{\pi/(4 \sqrt{C})\}$.
 
In the gapped region,
we extrapolate our numerical results to $\chi\rightarrow \infty$ with a fit in $1/\chi$ (for $\chi \geq70$), and consider only temperatures for which $\xi_\infty$ has a relative error of at most 3.5\% ($T \leq T_{\rm L} - \Delta T$, with $\Delta T\approx 0.02-0.03$).  From the extrapolated data we empirically find that $C\approx |t|/N$, and the overall prefactor $\xi_0(N)\approx \log(2)/N^{1.5}$.

\paragraph{Connection to a (2+1)D $\zn$ LGT.---}
\label{sec:app_LGT}
Here, we extend a known result from the deformed toric code model for $N=2$~\cite{castelnovo2008quantum} and we prove that Eq.~\eqref{eq:gs_LGT} is the ground state (GS) of a (2+1)D locally $\zn$-symmetric quantum Hamiltonian.
We first consider an infinite 2D lattice [see Fig.~\ref{fig:Fig5}(b)], with degrees of freedom located on the links. While in the main text we focus on one type of boundary conditions, the following proof is valid for general 2D topologies/boundary conditions.

We first define the $N$-dimensional ``clock'' operator \mbox{$Z=\sum_{n=0}^{N-1} \omega^n \ketbra{n}{n}$} and the cyclic ``shift'' operator \mbox{$X=\sum_{n=0}^{N-1} \ketbra{(n+1) \, \text{mod} \, N}{n}$}, with local basis states $\ket{n}$, $n=0, \dots, N-1$ and $\omega=e^{\frac{2\pi i}{N}}$. These operators fulfill the commutation relations $XZ=\omega^* Z X$ and $X^\dagger Z =\omega Z X^\dagger$, as well as $Z^N=X^N=1$. 
From the main text, we recall the Gauss' law $G_s(\{Z_l\})$ and plaquette operators $U_p(\{X_l\})$; from the above commutation relations, it is easy to check that $[U_p,G_s]=0$.
We further define the Abelian group $G=\{g\}$ with $g=\prod_p (U_p)^{n_p}$.

Consider now the Hamiltonian 
\begin{equation}
\textstyle
    H=\lambda_0 H_0 + \lambda_1 H_{\rm LGT},
\end{equation}
where $H_0=-\sum_s (G_s+G_s^\dagger)/2$ and $H_{\rm LGT}$ is given in Eq.~\eqref{eq:LGT_Ham}, with
the operators 
\begin{align}
    D_p^{(\prime)}(\beta)=e^{-\frac{\beta}{4}\left[d_N^{(\prime)*}\sum_{l=1}^2 Z_{p,l}+ d_N^{(\prime)} \sum_{l=3}^4 Z_{p,l} + {\rm H.c.}\right] },
\end{align}
where $d_N=1-\omega=d_N^{\prime *}$, and $\lambda_0, \lambda_1 >0$ are arbitrary couplings.
For $\lambda_0\rightarrow \infty$ we impose the Gauss' law constraint as $G_s=1, \forall s$, while in the limit $\beta\rightarrow 0$ $H$ reduces to the well-known $\zn$ toric code~\cite{horn1979hamiltonian}, up to an overall energy shift.

Our goal is to prove that the state
\begin{equation}
    \ket{\psi}=\sum_\alpha \frac{\psi_\alpha}{\sqrt{Z_\alpha}} \sum_{g\in G} e^{\frac{\beta}{2}\sum_i \frac{1}{2}(Z_i + Z_i ^\dagger)}g\ket{\psi_\alpha},
    \label{eq:app_gstopo}
\end{equation}
is the ground state of $H$, where $\{\ket{\psi_\alpha}\}$ form a minimal, non-unique set of reference configurations, from which under the action of $G$ the full Hilbert space basis can be generated, closely following the discussion in Ref.~\cite{castelnovo2008quantum} but now for $\mathbb{Z}_N$. Here, $\alpha$ labels the topological sectors of the $\zn$-toric code in non-trivial topologies, with
$\psi_\alpha$ the weight of the normalized wavefunction in each sector, i.e., $\sum_\alpha |\psi_\alpha|^2=1$, and $Z_\alpha$ the partition function, such that $\braket{\psi}{\psi}=1$.

We first show that $\ket{\psi}$ is a GS to $H_{\rm LGT}~=~\sum_p Q_p(\beta)$, with 
\begin{equation}
    \textstyle
    Q_p(\beta)=\frac{1}{2}(D_p(\beta)+D_p'(\beta) - U_p - U_p^\dagger).
\end{equation}
Using the commutation relations of $X$ and $Z$, and the fact that $U_p^{(\dagger)} g \in G$, direct computation shows that
$\ket{\psi}$ is annihilated by the operator $Q_p(\beta)$ and thus by $H_{\rm LGT}$, so it is an eigenstate with eigenvalue zero.
To show that $\ket{\psi}$ is also a GS, it is therefore sufficient to prove that $Q_p$ is lower-bounded by zero.
To do this, we first simplify the problem by considering conserved charges of $Q_p$, namely $Z_1^\dagger Z_2 = c_1, Z_2^\dagger Z_3^\dagger=c_2, Z_4 Z_1= c_4$, with $c_m=e^{2\pi i p_m/N}$ and $p_m=0, \dots, N-1$. 
It follows that $Q_p$ is block diagonal in the corresponding subsectors, with blocks $\tilde{Q}_p~=~(\tilde D_p+\tilde D_p' - X - X^\dagger)/2$, where
$\tilde{D}_p^{(\prime)}=\exp\{-\beta( d_N^{(\prime)*} c Z + {\rm H.c.})/4\}$,
with $c=1+c_1 + c_1 c_2 + c_4^*\in \mathbb{C}$, while $U_p^{(\dagger)}=X^{(\dagger)}$. Note that we dropped the label 1 in $Z_1^{(\dagger)}.$
We thus find that
\begin{equation}
\textstyle
    \tilde Q_p=\frac{1}{2}\sum_{n=0}^{N-1} \big[f_n \ketbra{n}{n} - \ketbra{n}{n-1} - \ketbra{n}{n+1}\big]
\end{equation}
is a real symmetric operator with 
\begin{equation}
\textstyle
    f_n=e^{-\frac{\beta}{4}[\omega^n d_N^* c + {\rm c.c.}]} +e^{-\frac{\beta}{4}[\omega^n d_N c + {\rm c.c.}]}\geq 0,
\end{equation}
and $\ket{N}\equiv\ket{0}, \ket{-1}\equiv\ket{N-1}$.
To show that $Q_p$ is positive-semidefinite, we now calculate $x^T \tilde Q_p x, \forall x=(x_0, \dots, x_{N-1})^T\in \mathbb{R}^N$.
Explicitly, 
\begin{equation}
\textstyle
    x^T \tilde Q_p x =\frac{1}{2}\sum_{n=0}^{N-1} \left[ x_n^2 f_n - (x_n x_{n+1} + x_{n-1} x_n)\right],
\end{equation}
with $x_N\equiv x_0$ and $x_{-1}\equiv x_{N-1}$.
Rewriting $f_n=g_n + 1/g_{n-1}$, with
\begin{equation}
\textstyle
    g_n=e^{-\frac{\beta}{4}(c \omega^n + {\rm c.c.})}/e^{-\frac{\beta}{4}(c \omega^{n+1}+ {\rm c.c.})} \geq0
\end{equation}
and shifting indices of selected terms $(n \rightarrow n+1)$ gives
\begin{equation}
    x^T \tilde Q_p x=\frac{1}{2}\sum_{n=0}^{N-1}\left( x_n \sqrt{g_n} - \frac{x_{n+1}}{\sqrt{g_n}}\right)^2 \geq 0 \;.
\end{equation}
To summarize, $\tilde Q_p$ and hence also $Q_p$ is lower-bounded by zero, which
implies that $\ket{\psi}$ is a GS of $H_{\rm LGT}$.

Finally, to show that $\ket{\psi}$ is also the GS of $H_0$, recall that for general topologies $\ket{\psi_\alpha}$ in Eq.~\eqref{eq:app_gstopo} are reference configurations in the different topological sectors of the $\zn$-toric code, which all fulfill $G_s=1, \forall s$. Since $[G_s^{(\dagger)}, g]=0 \; \forall g$, the application of $g$ cannot modify the sector, and $\ket{\psi}$ is a GS of $H_0$.
Thus, the state $\ket{\psi}$ is a ground state of any combination of the two Hamiltonians $H_0, H_{\rm LGT}$, and as such of $H$.

For an infinite system (with trivial boundaries), as discussed in the main text, the GS is unique. 
In this case, we choose a trivial reference state $\ket{\psi_{\alpha'}}=\otimes_l\ket{\rm 0}$ (i.e., $\psi_\alpha=\delta_{\alpha,\alpha'}$).
We next introduce the variable $e^{i\vartheta_p}$ on the plaquette $p$, and,
for a given configuration $\{ \vartheta_p\}$, the value of $Z_l$ on the link between two plaquettes $p, p'$ is uniquely determined as $Z_l \equiv e^{i(\vartheta_p-\vartheta_{p'})}$ (see also Ref.~\cite{horn1979hamiltonian}), and as such we rewrite $\sum_l (Z_l + Z_l^\dagger)/2=\sum_{\langle p, p' \rangle} \cos{(\vartheta_p - \vartheta_{p'})}$. 
By inserting these relations in Eq.~\eqref{eq:app_gstopo}, we obtain Eq.~\eqref{eq:gs_LGT}.
This state can be directly encoded in a 2D iPEPS tensor network, and it is indeed the state we use in our numerical simulations (see EM~\ref{sec:app_state_construction}).

\onecolumngrid
\section*{ SUPPLEMENTAL MATERIAL }

\setcounter{equation}{0}
\setcounter{figure}{0}
\setcounter{table}{0}
\renewcommand{\theequation}{S\arabic{equation}}
\renewcommand{\thefigure}{S\arabic{figure}}
\setlength\tabcolsep{10pt}
\setcounter{secnumdepth}{3}

\vspace{4mm}
\twocolumngrid
\section*{Rescaling comparison}\label{sec:SM_appA}
Here we show that our field-digitization scaling (FDS) analysis goes beyond a simple temperature rescaling, demonstrating that the former is valid around $T_L(N)$, while the latter is valid around $T=0$.

Consider the $N$-state clock model defined in Eq.~\eqref{eq:clock_Ham}.
In the limit of large $J/T$, i.e., at very small temperatures $T$,  expanding $\cos(\theta_i - \theta_j)\approx 1 -(\theta_i -\theta_j)^2 + \mathcal{O}((\theta_i-\theta_j)^4)$ where $\theta_i = 2\pi n_i/N$, 
gives the discrete Gaussian (DG) model 
\begin{equation}
    H_{\rm DG} = 2J \left(\frac{\pi}{N}\right)^2 \sum_{\langle i,j\rangle} (n_i - n_j)^2,
\end{equation}
when $n$ is extended to $n\in \mathbb{Z}$.
This identifies an effective coupling $K_{\rm DG}=2J/T \cdot (\pi/N)^2$, showing a phase transition at a given $K_{\rm DG, c}$~\cite{hasenbusch1997computing} and suggesting to relate different $N$-digitized models by a simple temperature rescaling $T/T_{\rm L}^{\rm DG}$, with $T_{\rm L}^{\rm DG}=K_{\rm DG, c}/(2J)\cdot (N/\pi)^2$.
The applicability of this approach relies on the
validity of the Taylor approximation, i.e., around $T=0$.
In contrast, the FDS analysis of the main text directly employs the effective Sine-Gordon field theory and therefore targets the vicinity of the critical point $T_{\rm L}(N)$. 

\begin{figure}[h!]
    \centering
    \includegraphics[width=\linewidth]{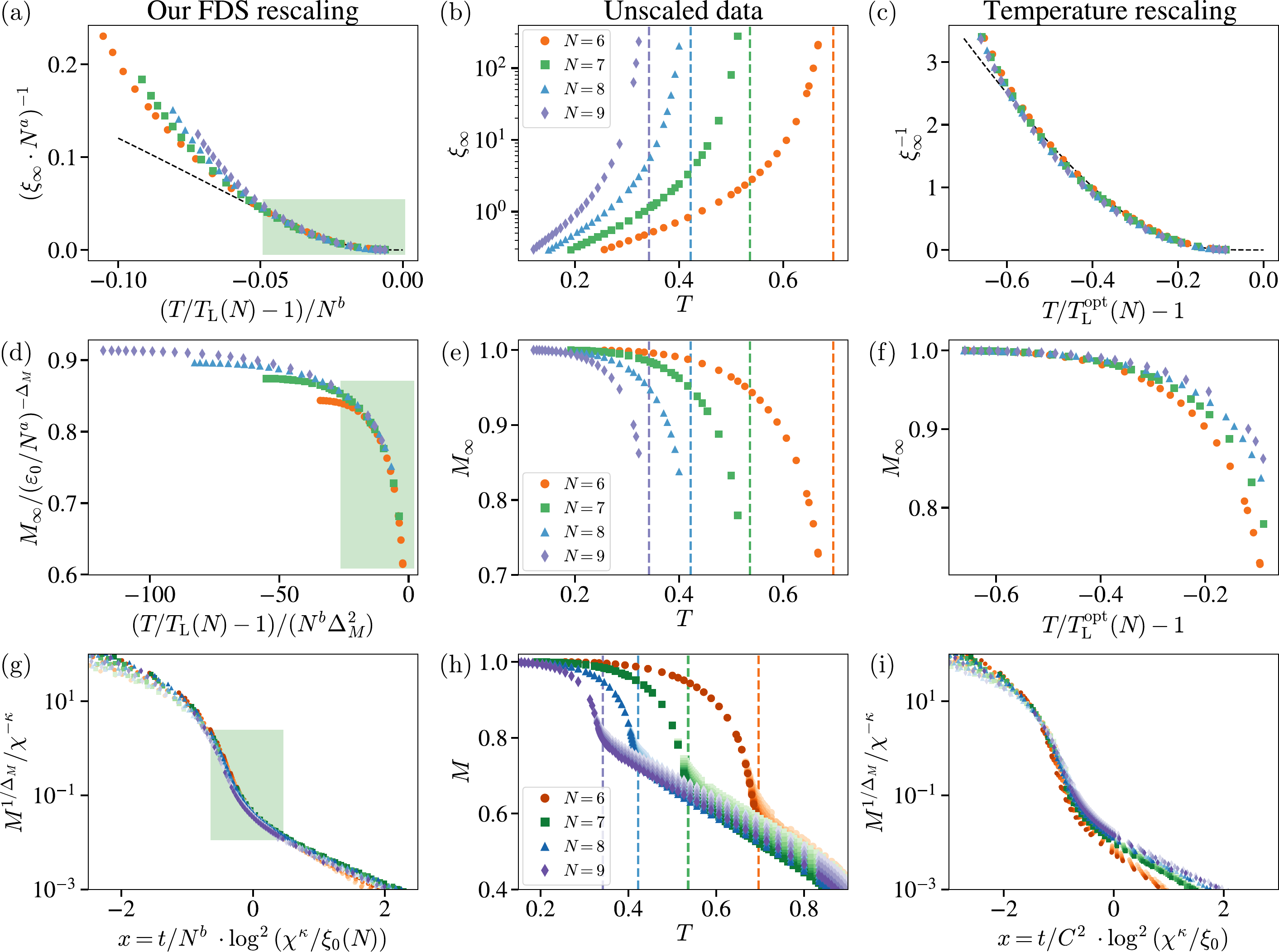}
    \caption{Collapses of observables in the ordered low-temperature gapped phase. We investigate the correlation length $\xi_\infty$ [(a)-(c)] and magnetization $M_\infty$ [(d)-(f)] extrapolated to the limit $\chi\rightarrow \infty$, as well as the magnetization $M$ at finite $\chi$ [(g)-(i)] for different rescalings of the axes. (Central column) Unscaled data. (Left column) Rescaling according to field-digitization scaling equations [Eqs.~\eqref{eq:corr_infty_num}-\eqref{eq:scalingAnsatz}]. The green shaded areas indicate the regime of validity of our rescaling. (Right column) Rescaled $x$-axis according to the prediction of the discrete Gaussian model. The critical temperatures are optimized as discussed in the SM. Throughout this figure, the numerical data are extrapolated to the limit $\chi\rightarrow \infty$ via a fit in $1/\chi$, with $\chi\geq 70$. For correlation length values $\xi$ smaller than a lattice sites some numerical problems emerge at large bond dimension $\chi$. Therefore, for temperatures at which $\xi(\chi=24)<0.6$, we consider only bond dimensions $\chi$ leading to correct results and perform the fit in $1/\chi$ with $\chi\geq 24$. }
    \label{fig:Fig1_SM}
\end{figure}

To make this distinction explicit, we compare two conceptually different rescalings
in Fig.~\ref{fig:Fig1_SM}, where
we plot the values of the correlation length $\xi_\infty$ [(a)-(c)] and magnetization $M_\infty$ [(d)-(f)] extrapolated to the limit $\chi\rightarrow \infty$ in the ordered phase $T<T_{\rm L}$, as well as the magnetization $M$ [(g)-(i)] for finite $\chi$.
In the central column of Fig.~\ref{fig:Fig1_SM} [plots (b),(e),(h)] we display the three quantities against the bare temperature $T$.
In the limit $T\rightarrow 0$, the magnetization converges to $M_\infty \rightarrow 1$ independently of $N$, as expected deep in the ordered phase.
In the left column Fig.~\ref{fig:Fig1_SM} [plots (a),(d),(g)] we rescale both axes according to the FDS predictions Eqs.~\eqref{eq:corr_infty_num}-\eqref{eq:scalingAnsatz} from the main text.
As anticipated, the curves collapse close to the critical point $T_{\rm L}$, both in the extrapolated case $\chi\rightarrow \infty$, as well as for finite $\chi$.
In the right column of Fig.~\ref{fig:Fig1_SM} [plots (c),(f),(i)], we instead rescale the $x$-axis according to the discrete Gaussian model prediction $( T/T_{\rm L} -1 )$. Additionally, we numerically optimize the critical temperatures $T_{\rm L}^{\rm opt} (N)$ by minimizing the $\chi^2_\mathrm{fit}$ value with $\chi^2_\mathrm{fit}=\sum_{n, i}(\log\xi_{n,i}-\log\xi_f(T_i))^2$. Here, $\xi_{n,i}=\xi_\infty(N_n, T_i)$ is the extrapolated value of the correlation length at a given $N_n$ and $T_i$ value, while $\xi_f(T)=\xi_0 \exp(C/\sqrt{|t|})$ is the standard functional form of the correlation length close to a BKT transition, with $t=T/T_\mathrm{L}(N)-1$ the reduced temperature. Following the derivation of the discrete Gaussian model, we assume that the whole $N$-dependence is captured by the critical temperature, while $\xi_0$ and $C$ are $N$-independent. From the optimization we obtain a value $\chi^2_\mathrm{fit}=0.536$, and the critical temperatures deviate by at most $6\%$ from the values in Ref.~\cite{li2022tensornetwork}.
For this numerically optimized temperature rescaling, the correlation length curves collapse in the whole gapped region, while the magnetization curves can be seen to collapse in the limit $T\rightarrow 0$. 
To investigate the magnetization $M$ at finite $N$ and $\chi$, in analogy with Eq.~\eqref{eq:scalingAnsatz}, we rescale the $x$-axis according to the predicted functional form $\xi(T)=\xi_0 \exp(C/\sqrt{|t|})$; this rescaling cannot fully account for the combined $N$- and $\chi$-dependence close to the critical point [see Fig.~\ref{fig:Fig1_SM}(i)]. For all the three observables, we observe the same qualitative behavior upon using literature values for the critical temperatures, as well as the naive prediction $T_{\rm L}^{\rm DG}$.

We conclude that at finite $N$ the two rescaling strategies indeed apply in two different regions and therefore probe distinct properties of the investigated models.

\onecolumngrid

\end{document}